\documentclass[twocolumn]{revtex4}
\usepackage[dvips]{graphicx}
\usepackage{bm}
\usepackage{amsmath}
\usepackage{lipsum}

\begin{document}

\title{Spin- and valley-dependent Goos-H\"{a}nchen effect in silicene and gapped graphene structures}

\author{E.S.~Azarova}
\author{G.M.~Maksimova}
\altaffiliation[Corresponding author. ]{Tel.: +7 831 4623304; \\
E-mail address: maksimova.galina@mail.ru (G.M. Maksimova)}

\address{Department of Theoretical Physics, University of Nizhny Novgorod, 23 Gagarin Avenue, 603950 Nizhny Novgorod, Russian Federation}

\begin{abstract}

We investigate the Goos-H\"{a}nchen shift for ballistic electrons (i) reflected from a step-like inhomogeneity of the potential energy and (or) effective mass, and (ii) transmitted through a ferromagnetic barrier region in monolayer silicene or gapped graphene. For the electrons reflected from a single interface we found that the Goos-H\"{a}nchen shift is valley-polarized for gapped graphene structure, and valley- and spin-polarized for silicene due to the spin-valley coupling. Incontrast, for example, to gapless graphene the lateral beam shift in gapped structures occurs not only in the case of total, but also of partial, reflection, i.e. at the angles smaller than the critical angle of total reflection. We have also demonstrated that the valley- and spin-polarized displacement of the electron beam, transmitted through a ferromagnetic silicene barrier, resonantly depends on the barrier width. The resonant values of the displacement can be controlled by adjusting the electric potential, the external perpendicular electric field, and the exchange field induced by an insulating ferromagnetic substrate.

\end{abstract}

\maketitle

\section*{INTRODUCTION}

It is well-known that a light beam incident on an interface of two dielectric media undergoes a lateral displacement under the condition of total internal reflection. This phenomenon was first observed in the experiment by Goos and H\"{a}nchen~\cite{GH}
, and named the Goos-H\"{a}nchen (GH) shift. Theoretical explanation of the effect was given by Artmann who used the stationary phase method~\cite{Art}
.  During the last decades, analogies of the spatial GH shift have been widely studied in acoustics, electronics, atomic optics, and particle physics. Interest in this phenomenon does not weaken up to now~\cite{Ign, Haan, Chen1, Duan, Savch, Zhou, Chen2, Yuan, Been, Lu, Song1}
. In particular, many works have been devoted to the electronic analog of the GH effect in semiconductor and graphene-based nanostructures with like-Dirac Hamiltonian (see Ref.~\cite{Chen3} 
 and references therein for review). The characteristic value of the GH displacement for Dirac electrons at a total reflection is of the order of a Fermi wavelength $\lambda_F$. Therefore, detecting the displacement is not a simple problem. However, as has been shown, the lateral shift for the electron beam tunneling through a single barrier structure can be enhanced by the transmission resonances when the incidence angle is less than the critical angle of total reflection~\cite{Lu, Chen4, Chen5}
 . Even greater effect arises in double-barrier structures (DBS). The GH effect of Dirac fermions in graphene DBSs was investigated by Song \textit{et al.}~\cite{Song1}
 . The authors found that at certain parameters the magnitude of the shift for transmitted electron beam increases dramatically and can reach values of the order of $1000$ $\lambda_F$ which are much greater than the maximum magnitude in the corresponding single barrier structure~\cite{Chen5}
 . It is remarkable that such giant GH shifts occur within the transmission gap and, as the authors suggest, their appearance is due to the quasibound states in the DBS. Such states are formed by the evanescent waves in the barriers. Indeed, as shown in Ref.~\cite{Song1}
 , significant difference in the phase shifts between the central and adjacent plane waves in the wave packets arises when the parameters of central plane wave correspond to the quasibound state. The ability of a control of the GH shift by electric and magnetic barriers or by strain-induced pseudo-magnetic fields can be used for design variety of devices such as spin beam splitter~\cite{Lu, Zhang1, Song2} 
 or valley beam splitter in graphene~\cite{Zhai, Zhang2, Wang}
 . However, the use of graphene in various device applications strikes on a strong restriction consisting in a lack of the energy gap in the electronic spectrum. This problem does not arise for some other promising two-dimensional materials, such as h-BN, dichalcogenides of transition metals, silicene and germanene having a honeycomb structure in $xy$-plane similarly to graphene. Silicene has been predicted by Takeda and Shiraishi~\cite{Takeda} 
and investigated in detail in Ref.~\cite{Guzm}
. There are two significant distinctions between silicene and graphene. First, it is a strong spin-orbit coupling (SOC) resulting in a band gap $\Delta_{so}$ in silicene spectrum, which is about $1.55-7.9$~ meV~\cite{Liu1, Liu2}
. Second, this is a buckled structure with two different sublattice planes separated by the distance $2l\approx0.046$~nm. This, in turn, makes it possible to control the energy gap in silicene by applying an external perpendicular electric field~\cite{Ni, Drum, Ezawa1, Cai}
. Germanene has similar properties.

In this paper we investigate the GH shift for Dirac fermions in silicene both at the total reflection and transmission through a barrier in the presence of inhomogeneous electric field, tunable potential $V$ and exchange field $h$. The results also can be applied to graphene structures, including the gapped graphene modification.

The effective low-energy Hamiltonian in silicene for Dirac electrons with spin $s$ $(s=\pm 1)$ in the vicinity of the $K$ $(\eta=1)$ and $K'$ $(\eta=-1)$ points is
 \begin{eqnarray}\label{eq1}
\hat{H}=\hbar\upsilon_F\left(k_x\tau_x{-}\eta k_y\tau_y\right){+}\Delta(x)\tau_z{+}V(x)I{-}h(x)sI.
\end{eqnarray}
Here $\upsilon_F\approx 0.5\times 10^6$ m/s is the Fermi velocity, $\tau_{x,y,z}$ are the Pauli matrices in the sublattice space, $I$ is the identity matrix, $V(x)$ is the barrier potential induced by the gate voltage. The second term in Eq.~(\ref{eq1}), $\Delta(x)=elE_z(x)-\eta s\lambda_{so}$ depending on the valley index and spin, describes the bandgap caused by the intrinsic SOC with strength $\lambda_{so}\approx3.9$~meV which is controlled by the external perpendicular electric field $E_z(x)$. The parameter $2l\approx0.046$~nm. For germanene $\lambda_{so}\approx43$~meV and $2l\approx0.066$~nm~\cite{Liu2}
. The exchange field $h(x)$ is assumed to be originated from the insulating ferromagnetic substrate. In particular, it has been analyzed theoretically if ferromagnetic insulator EuO is capable of creating spin splitting in graphene of the order of $5$ meV~\cite{Haug}
. In Ref.~\cite{Yang} 
Yang \textit{et al.} reported on the first-principles calculations of magnetic properties in graphene caused by the interaction of graphene with nearby europium oxide. Specifically, they found that large exchange-splitting band gap of about $30$ meV appears in the Dirac point. The influence of exchange field on spin and valley transport through arrays of silicene barriers has been studied in several works~\cite{Yok, Sood, Niu, Varg, Miss}
. Note, that in contrast to Refs~\cite{Sood, Ezawa2} 
we consider exchange fields which are the same for both sublattices. In the calculations we use the value $h = 3$ meV.

\section{Goos-H\"{a}nchen shift for the electron beam reflected from a single interface}

\begin{figure*}[t] \centering
\includegraphics*[scale=1.2]{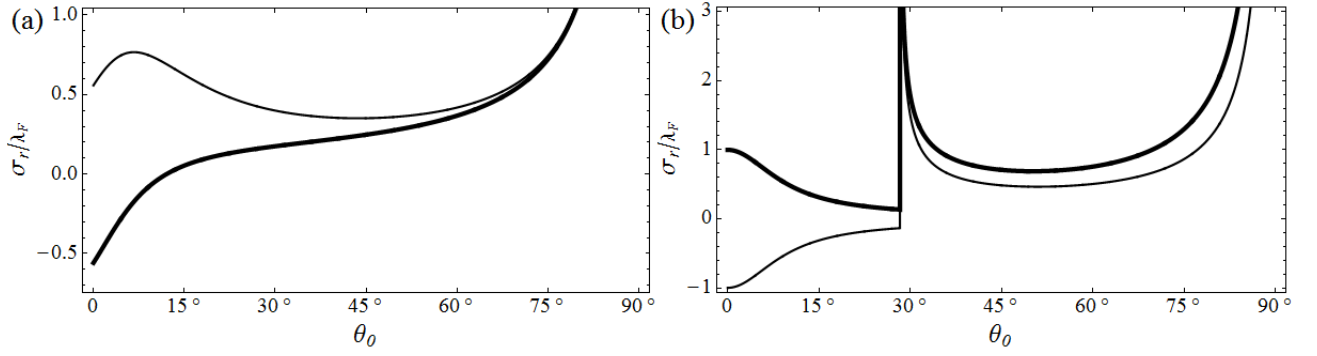}
\caption{The angular dependence of the GH shift in graphene for electrons belonging to $K$-valley ($\eta=1$, thick line) and $K'$-valley ($\eta=-1$, thin line) et (a) $E=100$~meV, $V=80$~meV, $\Delta_1=\Delta_2=26.5$~meV, and (b) $E=150$~meV, $V=80$~meV, $\Delta_1=26.5$~meV, $\Delta_2=0$.} \label{fig1}
\end{figure*}
Suppose that the sample is non-ferromagnetic, i.e. $h\!=\!0$ everywhere. The wave functions in the two different regions I (at $x<0$) and II (at $x>0$) can be written in terms of incident, reflected and evanescent waves with the incidence angle $\theta$. In the region I, where $V(x)=0$ and $\Delta(x)=\Delta_1=elE_{z1}-\eta s\lambda_{so}$, we have
\begin{eqnarray}\label{eq2}
\psi_I(x,y){=}e^{ik_xx{+}ik_yy}
\!\!
\begin{pmatrix}
1 \\                
Ce^{i\alpha_{in}}   
\end{pmatrix}
\!\!{+}re^{{-}ik_xx{+}ik_yy}
\!\!
\begin{pmatrix}
1 \\                
Ce^{i\alpha_r}      
\end{pmatrix}
\!{,}
\end{eqnarray}
where $k_x\!=\!k\cos\theta$, $k_y\!=\!k\sin\theta$ are the components of the electron wave vector with $k\!=\!\sqrt{E^2-\Delta_1^2}$, $\alpha_{in}\!=\!-\eta\theta$, $\alpha_r=\eta\theta+\pi$, $C=\sqrt{(E-\Delta_1)/(E+\Delta_1)}$ and $r$ is the reflection amplitude. When the angle $\theta$ exceeds the critical angle for total reflection
\begin{eqnarray}\label{eq3}
\theta_c=\arcsin\sqrt{\frac{\left(E-V\right)^2-\Delta_2^2}{E^2-\Delta_1^2}},
\end{eqnarray}
the evanescent solution in the second region (in which $V(x)\!=\!V$ and $\Delta(x)\!=\Delta_2=\!elE_{z2}\!-\!\eta s\lambda_{so}$) can be written as
\begin{eqnarray}\label{eq4}
\psi_{II}(x,y)=te^{-\kappa x+ik_yy}
\begin{pmatrix}
1 \\                
\frac{i\left(\kappa-\eta k_y\right)}{E-V+\Delta_2}   
\end{pmatrix}
,
\end{eqnarray}
where $\kappa=\sqrt{k_y^2+\Delta_2^2-(E-V)^2}$, and $t$ is the transmission amplitude. Hereafter $E$, $V$ and $\Delta_{1,2}$ are ``measured" in units of $\hbar\upsilon_F$. Matching the wave functions at the interface $x = 0$, we obtain the coefficients $t$ and $r$, where
\begin{eqnarray}\label{eq5}
\begin{gathered}
r=e^{-i\psi},\\
\psi=2\tan^{{-}1}\!\!\left(\frac{\eta k_y(E{-}V{+}\Delta_2)+\!\left(\kappa-\eta k_y\right)(E{+}\Delta_1)}{k_x(E{-}V{+}\Delta_2)}\right).
\end{gathered}
\end{eqnarray}
To find the GH shift we use the stationary phase method discussed in detail with regard to graphene in a number of works (see, e.g., Refs~\cite{Been, Song1, Chen3, Chen5}
).  However, due to the presence of the gap, the expression for the GH displacement of the reflected beam is not equivalent to the one obtained earlier for a gapless graphene. Assume that the wave functions of the incident and reflected beams have the form
\begin{eqnarray}\label{eq6}
\psi_{in}(x,y)\!=\!\int_{-\infty}^{\infty} f(k_y\!-\!k_{y0})e^{ik_xx+ik_yy}
\begin{pmatrix}
1\\
Ce^{i\alpha_{in}}
\end{pmatrix} dk_y,
\end{eqnarray}
\begin{eqnarray}\label{eq7}
\psi_{r}(x,y)\!=\!\int_{-\infty}^{\infty} rf(k_y\!-\!k_{y0})e^{-ik_xx+ik_yy}
\begin{pmatrix}
1\\
Ce^{i\alpha_{r}}
\end{pmatrix} dk_y,
\end{eqnarray}
where the quantities $C$, $\alpha_{in}$, $\alpha_r$ and $r$ are defined by Eqs~(\ref{eq2}) and (\ref{eq5}), and $f(k_y\!-\!k_{y0})$ is the angular spectrum distribution assumed to be of Gaussian shape around the central wave vector $k_{y0}$. For a well-collimated beam the position of the maximum of the upper or lower component of the wave function is determined by the requirement that the phase of this component be extremal as a function of $k_y$ at $k_y\!=\!k_{y0}$. Thus, at the interface $x\!=\!0$ the location of the upper ($y^+$) and lower ($y^-$) components of the incident wave function are: $y_{in}^+=0$, $y_{in}^-\!=\!-\partial\alpha_{in}/\partial k_{y0}$. Similarly, for the reflected wave packet: $y_r^+=\partial\psi/\partial k_{y0}$ and $y_r^-=\partial(\psi-\alpha_r)/\partial k_{y0}$, where ($-\psi$) is the phase of the reflection amplitude (Eq.~(\ref{eq5})). The shifts of the upper and lower components, respectively, are given by
\begin{eqnarray}\label{eq8}
\begin{gathered}
\sigma^+=y_r^+-y_{in}^+=\frac{\partial\psi}{\partial k_{y0}},\\
\sigma^-=y_r^--y_{in}^-=\frac{\partial}{\partial k_{y0}}\left(\psi+\alpha_{in}-\alpha_r\right).
\end{gathered}
\end{eqnarray}
The average GH displacement $\sigma_r$ is determined by both components. However, unlike a gapless graphene, their contributions to the wave packet are different: $1/\left(1+|C|^2\right)$ for the upper component, and $|C|^2/\left(1+|C|^2\right)$ for the lower one. Therefore, the expression for the GH shift in silicene (or gapped graphene) is determined by the formula:
\begin{eqnarray}\label{eq9}
\sigma_r=\frac{\sigma^++|C|^2\sigma^-}{1+|C|^2}.
\end{eqnarray}
For gapless graphene $|C|=1$ and $\sigma_r=\left(\sigma^++\sigma^-\right)/2$~\cite{Been}
. After straightforward calculations we obtain the GH shift in total reflection~(\ref{eq9}) in the following form:
\begin{widetext}
\begin{equation}\label{eq10}
\sigma_r(E,k_{y0})=\frac{{-}\eta\kappa(E{-}\Delta_1)\left[VF(E{+}\Delta_1){+}k_{y0}^2\Phi\right]
{+}k_{y0}^3(E{-}\Delta_1)\Phi{+}k_{y0}F\left[(E^2{-}\Delta_1^2)(V{+}\Delta_2){+}\Delta_1(\Delta_2^2{-}(E{-}V)^2)\right]}
{E\kappa k_x\left[F(E\Delta_2{-}E\Delta_1{+}V\Delta_1)+\!\left(k_{y0}^2{-}\eta\kappa k_{y0}\right)\Phi\right]},
\end{equation}
\end{widetext}
where $F=E-V+\Delta_2$, $\Phi=V+\Delta_1-\Delta_2$, $\Delta_1$ and $\Delta_2$ for given values of the electric field $E_{z1,2}$ in both half-spaces depend on the product of spin and valley indices. The values of $\kappa$ and $k_x$ are taken at $k_y=k_{y0}$. At appropriate choice of parameters the resulting expression~(\ref{eq10}) can be used for graphene. In the latter case $\Delta$ is the onsite potential difference between A and B sublattices, which does not depend on valley and spin indices, so that the GH displacements of the electrons belonging to the valleys $K$ and $K'$ satisfy the simple relation
\begin{equation}\label{eq11}
\sigma_r^{K'}\left(E,k_{y0}\right)=-\sigma_r^{K}\left(E,-k_{y0}\right).
\end{equation}
In particular, when $\Delta=0$ everywhere, we get the results of Ref.~\cite{Been} 
for gapless graphene. Fig.~\ref{fig1}(a) illustrates the dependence of the GH shift for gapped graphene at $\Delta_1=\Delta_2=\Delta$. Here, the parameters are chosen so that the condition for total internal reflection is satisfied at all angles $\theta_0$. For normal incidence, the displacement of electrons belonging to $K$ and $K'$ valleys differs only in sign: $\sigma_r(\theta_0=0)=-\eta\sqrt{E^2-\Delta^2}/(E\Delta)$. As the angle of incidence increases, this difference reduces, in fact, to zero. Our consideration shows that the existence of the gap in the graphene spectrum results in the valley-dependent GH effect. It is not so if the wave packet is reflected from the boundary of the p-n junction in gapless graphene~\cite{Been} 
or from the sharp heterojunction separating the gapless $(\Delta_1=0)$ and gapped $(\Delta_2\ne0,\text{ }V=0)$ graphene fractions. In the latter case, the GH shifts defined by the simple expression
\begin{equation}\label{eq12}
\sigma_r(E,\theta_0)=\frac{\hbar\upsilon_F\tan\theta_0}{\sqrt{\Delta_2^2-E^2\cos^2\theta_0}}.
\end{equation}
In addition, we found that in other gapped structures the lateral beam shift takes place not only in the case of total but also at partial reflection (when $\theta_0<\theta_c$) and has the form:
\begin{widetext}
\begin{equation}\label{eq13}
\sigma_r(E,\theta_0)=-\eta\frac{V(E^2-\Delta_1^2)(E\Delta_2+V\Delta_1-E\Delta_1)+k_y^2\Delta_1\left((\Delta_2-\Delta_1)^2-V^2\right)}
{Ek_x\left[(E\Delta_2+V\Delta_1-E\Delta_1)^2-k_y^2\left((\Delta_2-\Delta_1)^2-V^2\right)\right]}.
\end{equation}
\end{widetext}
Figure~\ref{fig1}(b) shows the angular dependence of the GH shifts for the valley-polarized electron beams reflected from the interface of gapped ($\Delta_1=26.5$~meV) and gapless ($\Delta_2=0$) graphene regions. Expression~(\ref{eq13}) describes this dependence within $0\le\theta_0<\theta_c\approx\pi/6$. At $\theta_0>\theta_c$, the displacements are computed using Eq.~(\ref{eq10}).

\begin{figure*}[t] \centering
\includegraphics*[scale=1.2]{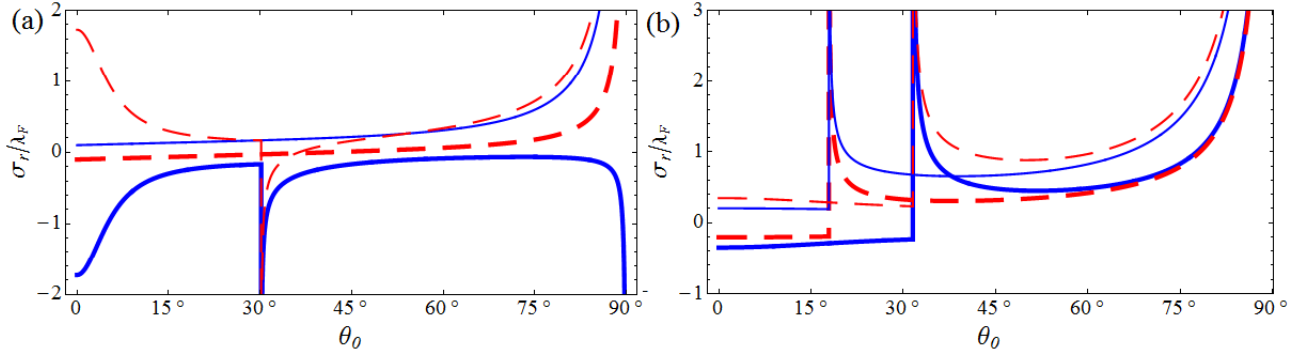}
\caption{The angular dependence of the GH shift in silicene for electrons belonging to $K$-valley (thick line) and $K'$-valley (thin line) with spin projections ``up" (blue solid lines) and ``down" (red dashed lines) at $\lambda_{so}=3.9$~meV, $V=11.7$~meV, $elE_{z1}=0$, $elE_{z2}=1.8\lambda_{so}$ for the energies $E=7.3$~meV (a) and $E=25$~meV (b).} \label{fig2}
\end{figure*}
In silicene the picture of the GH shifts $\sigma_r(E,\theta_0)$ becomes more complicated due to their dependence not only on the valley index but also on the spin direction (Fig.~\ref{fig2}). Therewith, the critical angle~(\ref{eq3}) depends on the product of the spin ($s$) and valley ($\eta$) indices. Thus, there are two critical angles $\theta_c$ corresponding to $\eta s=\pm1$ (Fig.~\ref{fig2}(b)). In particular, for a certain value of the product $\eta s$, we can so choose the parameters of the structure that the condition of total internal reflection will be satisfied at all angles $\theta_0$. Fig.~\ref{fig2}(b) illustrates such a case for $\eta s=-1$.

When reflecting from a single interface, the value of the lateral displacement (which can be positive or negative) is of the order of several Fermi wavelengths. However, in order to detect the valley-dependent (or spin-valley-dependent) GH shift, the difference of the displacements for $K$ and $K'$ valleys should be greater than the longitudinal width of the initial electron beam, which is about $100-1000$~Fermi wavelengths~\cite{Song1}
. A stronger effect, i.e. significant increase of the shift magnitude, can be achieved for the electronic beam passing through the single- or double-barrier structures~\cite{Song1, Chen5}
.

\section{Goos-H\"{a}nchen-like shift for Dirac fermions transmitted through a single barrier}

The lateral displacement of electron beams tunneling through a ferromagnetic barrier structure formed by an electrostatic potential $V$ and exchange field $h$ in silicene (or graphene) sheet is defined by the transmission amplitude $t$ of the plane-wave solution of the Dirac equation
\begin{equation}\label{eq14}
\psi_t(x,y)=te^{ik_x(x-d)+ik_yy}
\begin{pmatrix}
1\\
Ce^{i\alpha_{in}}
\end{pmatrix}.
\end{equation}
Here $t=|t|e^{-i\varphi}$ can be found from the relation~\cite{Az}
\begin{equation}\label{eq15}
\frac{1}{t}=\cos\beta+i\frac{E(V-sh)+\Delta_1\Delta_2-E^2+k_y^2}{k_xq_x}\sin\beta,
\end{equation}
where $q_x=\sqrt{(E-V+sh)^2-\Delta_2^2-k_y^2}$ and $k_x=\sqrt{E^2-\Delta_1^2-k_y^2}$ are the wave vectors inside and outside the barrier, respectively, $\beta=q_xd$, and $d$ is the width of the potential barrier. The gap parameters $\Delta_1$ and $\Delta_2$ are defined by the values of the electric field $E_{z1}$ and $E_{z2}$ in the corresponding regions (in silicene). In the case where the incident beam is well collimated around some transverse wave vector $k_{y0}$, the GH shift for the transmitted beam is $\sigma_t(E,k_{y0})=d\varphi/dk_{y0}$~\cite{Song1, Chen5}
. When the incidence angle $\theta_0$ is less than the critical angle $\theta_c$ for the total reflection (Eq.~(\ref{eq3})), we obtain from Eq.~(\ref{eq15}):
\begin{widetext}
\begin{equation}\label{eq16}
\sigma_t(E,k_{y0})=k_{y0}\frac{f\left(E,k_{y0}\right)\left[-k_xd(1+\tan^2\beta)+\frac{k_x^2+q_x^2}{k_xq_x}\tan\beta\right]+2\tan\beta}
{k_xq_x\left(1+f^2(E,k_{y0})\tan^2\beta\right)},
\end{equation}
\end{widetext}
where wave vectors $k_x$, $q_x$ are taken at $k_y=k_{y0}$, and
\begin{equation}\label{eq17}
f(E,k_y)=\frac{E(V-sh-E)+\Delta_1\Delta_2+k_y^2}{k_xq_x}.
\end{equation}
Note, that angles of incidence exceeding the critical value $\theta_c$, the lateral shift can be obtained from these equations by replacing $q_x\to i\kappa$, where $\kappa=\sqrt{k_y^2+\Delta_2^2-(E-V-hs)^2}$.
\begin{figure}[t] \centering
\includegraphics*[scale=0.6]{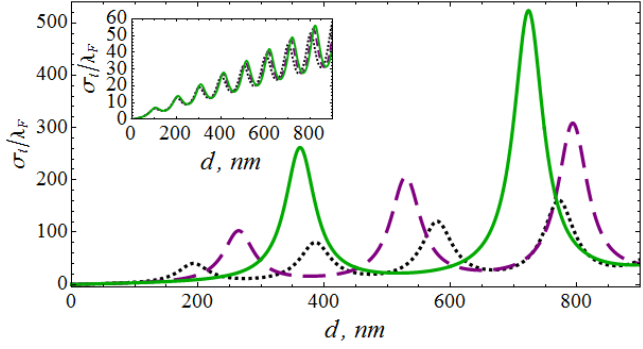}
\caption{GH shifts in transmission through non-ferromagnetic barrier in silicene as a function of the barrier width $d$ at $\eta s=-1$, $E=40$~meV, $V=7.46$~meV, $elE_{z1}=0$ and $\theta_0=50^{\circ}$. Solid (green), dashed (purple), and dotted (black) lines correspond to $elE_{z2}=1.8\lambda_{so},\text{ }1.7\lambda_{so}\text{ and }1.5\lambda_{so}$, respectively. Insert: the same as in the main figure, but for the spin-valley polarization $\eta s=1$.} \label{fig3}
\end{figure}

As follows from equations~(\ref{eq16}), (\ref{eq17}) the GH shift for the electrons, transmitted through the non-ferromagnetic barrier in graphene is the same for both valleys and both values of spin. In the absence of perpendicular electric field this result takes place also for silicene. However, applied electric field leads to a GH shift, depending on the spin-valley index $\eta s$ (Fig.~\ref{fig3}). In the case of propagating beam (i.e. at $\theta_0<\theta_c$) the dependence of the lateral displacement on the barrier width $d$ has a pronounced resonant character. The resonance widths $d_n$, at which the barrier is transparent, are determined from the relations $q_xd=\pi n,\text{ }n=1,2,\dots$. For such a barrier structure we immediately find from Eq.~(\ref{eq15}) the resonant values of the GH shifts
\begin{figure}[t] \centering
\includegraphics*[scale=0.6]{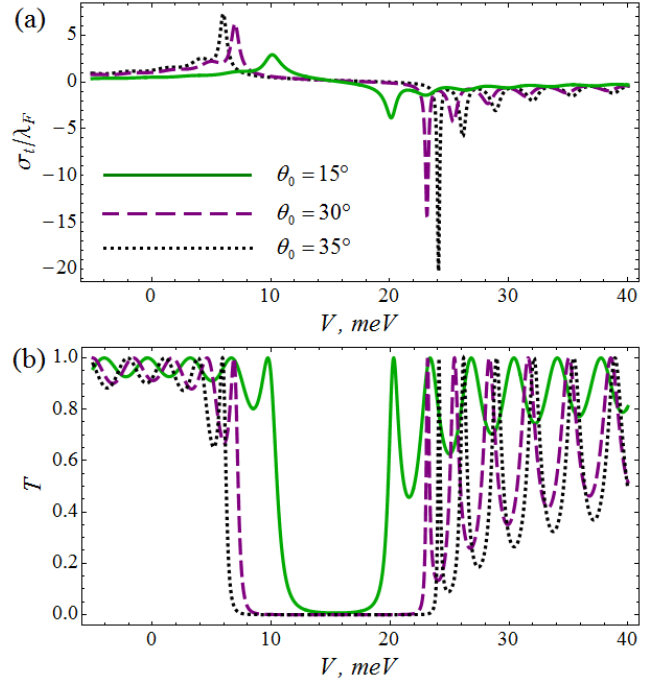}
\caption{(a) GH shifts and (b) transmission probability through non-ferromagnetic barrier in silicene as a function of the barrier hight $V$ at $E=15$ meV, $d=300$ nm, $elE_{z1}=0$, $elE_{z2}=\lambda_{so}$ and $\eta s=1$.} \label{fig4}
\end{figure}
\begin{equation}\label{eq18}
\sigma_t^n(E,\theta_0)=-\pi nk_yf(E,k_y)q_x^{-2},
\end{equation}
which can be modulated by the potential barrier height and the induced gaps $\Delta_1$ and $\Delta_2$. Thus, as follows from Eq. (\ref{eq18}), the increase of the gap $\Delta_1$ outside the barrier suppresses the effect, i.e. leads to a decrease in the magnitude of $\sigma_t$. At the same time, comparison of the curves in Fig.~\ref{fig3} corresponding to different values of $\Delta_2$ shows that the absolute value of the shift increases with increasing the gap parameter in the barrier which is consistent with the results of Ref.~\cite{Chen5}
. Thus, for spin-valley polarization $\eta s=1$, the electric field in the barrier region reduces the effective gap parameter $\Delta_2$ and, consequently, the value of $\sigma_t$ approximately by the order of magnitude compared to its value at $\eta s=-1$ (see insert in Fig.~\ref{fig3}).

In Fig.~\ref{fig4} we present (a) the GH shift and (b) transmission probability $T=|t|^2$ (Eq.~(\ref{eq15})) as function of the barrier height at different angles of incidence. We see that outside the transmission gap the GH shift $\sigma_t(V)$ exhibits a resonance structure. The resonance values of the potential are defined as $V_n^{\pm}=E\pm\sqrt{(\pi n/d)^2+\Delta_2^2+(k\sin\theta_0)^2}$ with $k=\sqrt{E^2-\Delta_1^2}$. Thus, for a given $n$ ($n = 1,2\dots$), there are two resonances located to the right ($V_n^+$) and left ($V_n^-$) from the transmission gap (Fig.~\ref{fig4}(b)), and corresponding to the Klein and classical tunneling respectively. Resonant displacement $\sigma_t^{n\pm}=\sigma_t(V_n^{\pm})$ is obtained from Eq.~(\ref{eq18}), and has the form
\begin{eqnarray}\label{eq19}
&&\sigma_t^{n\pm}=-\frac{d^3\tan\theta_0}{(\pi n)^2}\Bigl[\pm E\sqrt{(\pi n/d)^2+\Delta_2^2+(k\sin\theta_0)^2}+ \nonumber \\
&&\Delta_1\Delta_2+(k\sin\theta_0)^2 \Bigl].
\end{eqnarray}
\begin{figure*}[t] \centering
\includegraphics*[scale=1.2]{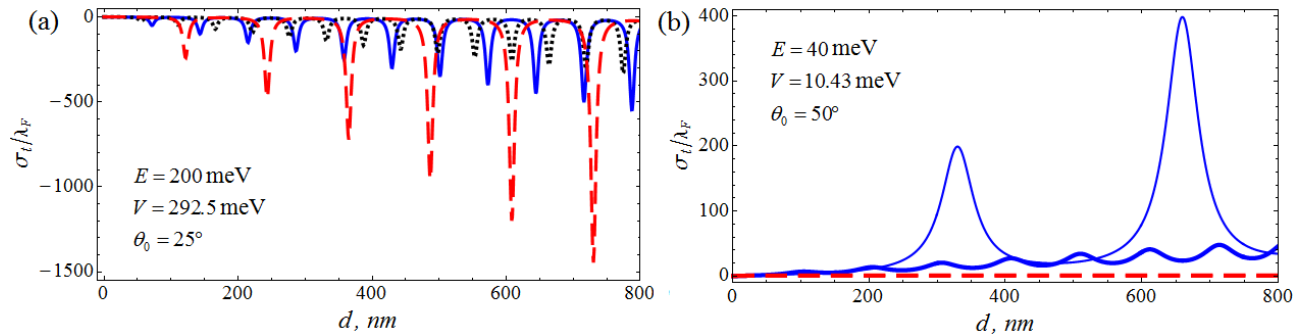}
\caption{GH shifts in transmission through a ferromagnetic barrier as a function of the barrier width $d$: (a) for graphene with $\Delta_1=\Delta_2=26.5$~meV and $h=0$ (blue solid line); $h=3$~meV, $s=1$ (red dashed line); $h=3$~meV, $s=-1$ (black dotted line), and (b) for silicene with $elE_{z1}=0$, $elE_{z2}=1.8\lambda_{so}$ and $h=3$~meV: thick and thin solid blue lines correspond to electrons with spin up belonging to $K$ and $K'$- valley, respectively. The lateral displacement for spin down polarization is much less (red dashed line) due to the evanescent character of the waves inside the barrier.} \label{fig5}
\end{figure*}
This expression clearly demonstrates that, (i) the magnitude of the displacement decreases rapidly with increasing resonance number $n$, and (ii) the GH shift is negative for Klein tunneling. In this case, increase of the gap value $\Delta_2$ results in the increase of the $n$-th resonance amplitude. Increase of the angle of incidence $\theta_0$ leads to the same effect (Fig.~\ref{fig4}(a)). Not so obvious, but can be shown, that in the classical case ($E > V$) lateral shifts become positive and, as follows from Eq.~(\ref{eq16}), at given $n$ the magnitude of Klein displacement $\sigma_t^{n+}$ is greater than $\sigma_t^{n-}$ (Fig.~\ref{fig4}(a)). Inside the transmission gap $E-\sqrt{\Delta_2^2+(k\sin\theta_0)^2}<V<E+\sqrt{\Delta_2^2+(k\sin\theta_0)^2}$ that corresponds to the evanescent solutions the GH shift is of the order of electron wavelength similarly to the case of the GH effect at a single graphene interface. We also examined modulation of the GH shift depending on energy $E$ of the incident beam. Corresponding dependence $\sigma_t(E)$ is similar to the one discussed above but the Klein and classical zones (and, accordingly, the signs of the GH shifts) are now interchanged.

For electrons transmitted through the ferromagnetic region ($h(x)=h$ at $0\le x\le d$) the height of the effective potential barrier $V_{eff}=V-sh$ turns out to be spin-dependent. This influences the transport properties of both silicene~\cite{Yok, Sood, Niu, Varg, Miss} 
and graphene~\cite{Haug}
. In particular, the GH displacement $\sigma_t(E,k_{y0})$ becomes spin-polarized for graphene (Fig.~\ref{fig5}(a)) and, due to the correlation between the valley and spin degrees of freedom, spin-and-valley-polarized for silicene (Fig.\ref{fig5}(b)). The GH shifts obtained above can be positive or negative depending on the ratio between the electron energy $E$ and barrier height $V_{eff}$. Thus, all curves in Fig.~\ref{fig5}(a) correspond to the case of Klein tunneling ($E<V_{eff}-\Delta_2$) through a ferromagnetic barrier region in gapped graphene. In this case $\sigma_t(E,k_{y0})$ is negative for both spin channels and, as follows from Eq.~(\ref{eq18}), the magnitude of the shift in the $n$-th resonance increases with decreasing the difference between $V_{eff}-\Delta_2$ and $E$. The GH shifts for classical tunneling ($E>V_{eff}+\Delta_2$) in silicene are shown in Fig.~\ref{fig5}(b) by solid lines for both valleys with $s=1$. In this case the barrier heights for electrons belonging to different valleys are equivalent, but the magnitudes of the gap inside the barrier differ: $\Delta_2(K)=0.8\lambda_{so}$, $\Delta_2(K')=2.8\lambda_{so}$. Accordingly, the electrons from the $K'$-valley have greater displacement than the $K$-polarized electrons. For the parameters used, electrons with the spin-down projections in both valleys are described by the evanescent waves inside the barrier which leads to the suppression the GH effect (red dashed line in Fig.~\ref{fig5}(b)).

\section*{SUMMARY}

We have investigated the GH shifts for Dirac fermions totally reflected from the profile of the potential energy and effective mass as well as transmitted through a ferromagnetic barrier region in silicene and gapped graphene. We have shown that the presence of the gap in the graphene spectrum results in the valley-dependent GH effect in total reflection. In silicene the lateral displacement of the reflected beam also depends on the spin direction due to the coupling between valley and spin degrees of freedom. It was also found that in gapped structures the lateral beam shift occurs not only in the case of total, but also in partial reflection (Eq.~(\ref{eq13})). For the electrons, transmitted through the barrier region the GH shift can be enhanced by the transmission resonances. The resonant values of the displacement depend on the incidence angle and the electron energy as well as on the structure characteristics such as the barrier height and gap. In particular, our results show that the GH shift increases (decreases) with increasing the gap inside (outside) the barrier. We have also demonstrated that for a normal/ferromagnetic/normal silicene junction the GH shift is valley- and spin-polarized. The obtained results can also be applicable to other two-dimensional hexagonal crystals, such as germanene or monolayers of MoS$_2$ and other group VI dichalcogenides, which have two inequivalent valleys and two inequivalent lattices.

\section*{ACKNOWLEDGMENTS}

This work was supported by the Russian Foundation for Basic Research (Grant No. $16-32-00712$-mol$\_$a). E.S.A. acknowledges support by the ``Dynasty'' Foundation.


\begin{thebibliography}{99}

\bibitem{GH}
F.~Goos, H.~H\"{a}nchen, Ann. Phys. 1 (1947) 333; 5 (1949) 251.

\bibitem{Art}
K.V.~Artmann, Ann. Phys. 2 (1948) 87.

\bibitem{Ign}
V.K.~Ignatovich, Phys. Lett. A 322 (2004) 36.

\bibitem{Haan}
V.-O.~de~Haan, J.~Plomp, T.M.~Rekveldt, W.H.~Kraan, Ad~A.~van~Well, R.M.~Dalgliesh, S.~Langridge, Phys. Rev. Lett. 104 (2010) 010401.

\bibitem{Chen1}
X.~Chen, C.-F.~Li, Phys. Rev. E 69 (2004) 066617.

\bibitem{Duan}
Z.~Duan, L.~Hu, X.X.~Xu, C.~Liu, Opt. Exp. 22 (2014) 17679.

\bibitem{Savch}
A.S.~Savchenko, A.S.~Tarasenko, S.V.~Tarasenko, V.G.~Shavrov, JETP Lett. 102:6 (2015) 380.

\bibitem{Zhou}
L.~Zhou, J.-L.~Qin, Z.~Lan, G.~Dong, W.~Zhang, Phys. Rev. A 91 (2015) 031603.

\bibitem{Chen2}
X.~Chen, X.-J.~Lu, Y.~Wang, C.-F.~Li, Phys. Rev. B 83 (2011) 195409.

\bibitem{Yuan}
L.~Yuan, L.-L.~Xiang, Y.-H.~Kong, M.-W.~Lu, Z.-J.~Lan, A.-H.~Zeng, Z.-Y.~Wang, Eur. Phys. J. B 85 (2012) 1434.

\bibitem{Been}
C.W.J.~Beenakker, R.A.~Sepkhanov, A.R.~Akhmerov, J.~Tworzyd{\l}o, Phys. Rev. Lett. 102 (2009) 146804.

\bibitem{Lu}
M.-W.~Lu, G.-L.~Zhang, S.-Y.~Chen, J. Appl. Phys. 112 (2012) 014309.

\bibitem{Song1}
Y.~Song, H.-C.~Wu, Y.~Guo, Appl. Phys. Lett. 100 (2012) 253116

\bibitem{Chen3}
X.~Chen, X.-J.~Lu, Y.~Ban, C.-F~Li, J. Opt. 15 (2013) 033001.

\bibitem{Chen4}
X.~Chen, C.-F.~Li, Y.~Ban, Phys. Lett. A 354 (2006) 161.

\bibitem{Chen5}
X.~Chen, J.-W.~Tao, Y.~Ban, Eur. Phys. J. B 79 (2011) 203.

\bibitem{Zhang1}
Q.~Zhang and K.S. Chan, Appl. Phys. Lett. 105 (2014) 212408.

\bibitem{Song2}
Y.~Song and G. Dai, Appl. Phys. Lett. 106 (2015) 223104.

\bibitem{Zhai}
F.~Zhai, Y.-L.~Ma, K.~Chang, New J. Phys. 13 (2011) 083029.

\bibitem{Zhang2}
Q.~Zhao and K.S.~Chan, RSC Advances 5 (2015) 8371.

\bibitem{Wang}
J.~Wang, M.~Long, W.-S.~Zhao, Y.~Hu, G.~Wang, and K.S.~Chan, J. Phys.: Condens. Matter 28 (2016) 285302.

\bibitem{Takeda}
K.~Takeda, K.~Shiraishi, Phys. Rev. B 50 (1994) 14916.

\bibitem{Guzm}
G.G.~Guzm{\'a}n-Verri, L.C.~Lew Yan Voon, Phys. Rev. B 76 (2007) 075131.

\bibitem{Liu1}
C.-C~Liu, W.~Feng, Y.~Yao, Phys. Rev. Lett. 107 (2011) 076802.

\bibitem{Liu2}
C.-C.~Liu, H.~Jiang, Y.~Yao, Phys. Rev. B 84 (2011) 195430.

\bibitem{Ni}
Z.~Ni, Q.~Liu, K.~Tang, J.~Zheng, J.~Zhou, R.~Qin, Z.~Gao, D.~Yu, J.~Lu, Nano Lett. 12 (2012) 113.

\bibitem{Drum}
N.D.~Drummond, V.~Z{\'o}lyomi, V.I.~Fal’ko, Phys. Rev. B 85 (2012) 075423.

\bibitem{Ezawa1}
M.~Ezawa, New J. Phys. 14 (2012) 033003.

\bibitem{Cai}
Y.~Cai, C.-P.~Chuu, C.M.~Wei, M.Y.~Chou, Phys. Rev. B 88 (2013) 245408.

\bibitem{Haug}
H.~Haugen, D.~Huertas-Hernando, A.~Brataas, Phys. Rev. B 77 (2008) 115406.

\bibitem{Yang}
H.X.~Yang, A.~Hallal, D.~Terrade, X.~Wainta, S.~Roche, and M.~Chshiev, Phys. Rev. Lett. 110 (2013) 046603.

\bibitem{Yok}
T. Yokoyama, Phys. Rev. B 87 (2013) 241409.

\bibitem{Sood}
B. Soodchomshom, J. Appl. Phys. 115 (2014) 023706.

\bibitem{Niu}
Z.P.~Niu and S. Dong, Appl. Phys. Lett. 104 (2014) 202401.

\bibitem{Varg}
V.~Vargiamidis and P.~Vasilopoulos, Appl. Phys. Lett. 105 (2014) 223105; J. Appl. Phys. 117 (2015) 094305.

\bibitem{Miss}
N.~Missault, P.~Vasilopoulos, V.~Vargiamidis, F.M.~Peeters, B.~Van~Duppen, Phys. Rev. B 92 (2015) 195423.

\bibitem{Ezawa2}
M. Ezawa, Phys. Rev. B 87 (2013) 155415.

\bibitem{Az}
E.S.~Azarova, G.M.~Maksimova, Physica E 74 (2015) 1.

\end{thebibliography}
\end{document}